\documentclass[aps,prl,superscriptaddress,showpacs,twocolumn,
]{revtex4-1}
\usepackage{graphicx}
\usepackage{epstopdf}
\usepackage{natbib}

\begin{document}


\title{Magnon-mediated thermal transport in antiferromagnets: the link to momentum-resolved magnon lifetime data}


\author{S.P. Bayrakci}
\email[]{bayrakci@fkf.mpg.de}
\affiliation{Max-Planck-Institut f\"ur Festk\"orperforschung, Heisenbergstrasse 1, D-70569 Stuttgart, Germany}
\author{B. Keimer}
\affiliation{Max-Planck-Institut f\"ur Festk\"orperforschung, Heisenbergstrasse 1, D-70569 Stuttgart, Germany}
\author{D.A. Tennant}
\affiliation{Helmholtz-Zentrum Berlin f\"ur Materialien und Energie, Hahn-Meitner-Platz 1, D-14109 Berlin, Germany}
\affiliation{Institut f\"ur Festk\"orperphysik, Technische Universit\"at Berlin, Hardenbergstrasse 36, D-10623 Berlin, Germany}


\date{\today}

\begin{abstract}
Transport currents in solids decay through collisions of quasiparticles with each other and with defects or boundaries. Since information about collisional lifetimes is difficult to obtain, most calculations of transport properties rely on parameters that are not known independently. Here, we use magnon lifetime data for the two-dimensional antiferromagnet Rb$_{2}$MnF$_{4}$ to calculate the magnon-mediated thermal conductivity without any adjustable parameters, thereby quantifying the influence of scattering from domain boundaries on transport.  Related strategies have the potential to enhance our understanding of thermal transport by electronic and phononic quasiparticles greatly.
\end{abstract}


\maketitle


Thermal transport phenomena are of great importance for many areas of science and technology, ranging from correlated electron systems\cite{foo,doiron} to the design of nanostructured materials for solar energy harvesting\cite{hochbaum}.  A prerequisite for progress towards nanoscale and quantum devices is thorough understanding of the effects of thermal noise and quantum coherence and their relation to size, dimension, and coupling to the environment.  In quantum materials, thermal effects are of fundamental importance, as they affect the transport, storage, and processing of energy and information. 

It is widely assumed that the thermal conductivity mediated by phonons in simple insulators and undoped semiconductors is well understood.  In fact, even in the most advanced investigations of the thermal conductivity of elemental semiconductors reported to date, multiple adjustable parameters describing the collisional rates of longitudinal and transverse phonons had to be introduced to fit the experimental data\cite{asen-palmer}.  Recent experiments on two-dimensional (2D) antiferromagnetic insulators\cite{hess,hofmann,li,berggold} have allowed the determination of the thermal conductivity $\kappa_{\mathrm{mag}}$ mediated by magnons.  Their theoretical description requires fewer parameters than that of phonons because only two degenerate (or nearly-degenerate) magnon branches are present.  However, the resulting data were analyzed only at low temperatures, where $\kappa_{\mathrm{mag}}$ is limited by defects, and the defect density could not be determined independently.  Clearly, a more complete and quantitative understanding of thermal conduction in such model systems is required as a basis for theories of complex materials that are of current scientific and technological interest.  

We present an approach to this problem that establishes a quantitative link over a broad range of temperatures between the macroscopic property of the thermal transport and the microscopic processes that underlie it, without the need for adjustable parameters.  This approach is made possible by the advent of a new neutron spin-echo technique by means of which excitation linewidths can be measured with high resolution in energy and wavevector over the full Brillouin zone\cite{kellerNRSE}, combined essentially with a theoretical description of these linewidths that is quantitatively accurate.  Very recently, magnon linewidths in 2D and 3D antiferromagnets were measured over wide ranges of temperature using this technique, and a comprehensive theoretical description that includes numerical evaluation of 4-magnon (two-in/two-out) scattering processes, combined with a small contribution from boundary scattering, was developed\cite{bayrakcipreprint}.

In general, experimental identification of $\kappa_{\mathrm{mag}}$ is easier in magnetic systems of low dimensionality.  Since the phononic contribution to the thermal conductivity $\kappa$ is roughly isotropic, subtracting $\kappa$ measured in a direction of negligible magnetic dispersion from that measured along a dispersive direction yields an estimate of $\kappa_{\mathrm{mag}}$ associated with the dispersive magnons.  In a number of quasi-1D and 2D magnetic systems, $\kappa_{\mathrm{mag}}$ appears as a broad peak that is well separated in temperature from the phononic peak\cite{kudo,sologubenko,hess}.  As the spinon excitations in quasi-1D compounds occupy a continuum in energy, measurement of their linewidths is inherently more difficult, and corresponding theoretical descriptions are less well developed.  We will therefore consider only higher-dimensional magnetic systems in our analysis.

The Boltzmann transport equation describes the irreversible approach of a perturbed dilute gas to equilibrium.  It can be used to quantify heat flow in the presence of a temperature gradient.  In order for a perturbative interaction in a quantum system to allow the system to approach equilibrium in the manner described by this equation, a large phase space of scattered states must exist, and the coupling between scattered states must be weak enough that there is essentially negligible correlation between them\cite{vanHove}.  The scattering events are then elastic, with probabilities given by Fermi's Golden Rule.  Magnon-magnon and phonon-phonon interactions at low temperatures in higher-dimensional systems satisfy the above conditions\cite{vanHove}.  Exponential relaxation of such interacting quasiparticles to equilibrium yields Lorentzian excitation lineshapes.  A comprehensive description of these relaxation rates permits quantification of the heat current and the drift distribution of the quasiparticles.

The heat flow per unit area $\mathbf{Q} = \kappa\nabla\!\mathit{T}$ due to thermally-excited collective modes with thermal conductivity $\kappa$ in the presence of a thermal gradient $\nabla\!\mathit{T}$ can be calculated from the deviations $\delta\mathit{n}_{\mathit{k}}$ of the mode occupation numbers from equilibrium by means of the relation
\begin{equation}
\kappa = -\frac{2}{\mathit{V}_{0}|\nabla\!\mathit{T}|^{2}}\sum_{k}\hbar\omega_{\mathit{k}}\,\delta\mathit{n}_{\mathit{k}}\,\mathbf{c}_{\mathit{k}}\!\cdot\!\nabla\!\mathit{T},
\end{equation}
\noindent where $\hbar\omega_{\mathit{k}}$ is the excitation energy and $\mathbf{c}_{\mathit{k}}$ its group velocity, and $\mathit{V}_{0}$ is the volume of the unit cell.  Values of $\delta\mathit{n}_{\mathit{k}}$ can in turn be calculated from the excitation lifetimes by means of the approximate solution of the Boltzmann transport equation proposed by Callaway\cite{callaway}.  Callaway considered phonons with linear spin-wave dispersions.  In the following, we proceed as in his calculation, generalizing it to apply to arbitrary spin-wave dispersions.  We treat magnons, rather than phonons; the factor of 2 in the above expression for $\kappa$ originates from the aforementioned twofold degeneracy of the spin waves.

According to the Boltzmann equation, the mode occupation numbers evolve in response to both 1) the heat flowing into a volume element over a local thermal gradient and 2) changes in mode occupation numbers due to collisions.  In steady-state heat flow, the mode occupation numbers remain constant, and thus the change in the mode distribution due to collisions is balanced by the variation in mode occupation over the thermal gradient.  This is expressed as
\begin{equation}
\frac{\partial\mathit{n}_{\mathit{k}}}{\partial\mathit{t}} = \mathbf{c}_{\mathit{k}}\!\cdot\!\nabla\!\mathit{T}\frac{\mathit{d}\mathit{n}_{\mathit{k}}}{\mathit{dT}},
\end{equation}
\noindent where $\mathit{n}_{\mathit{k}}$ is  a mode occupation number and $\partial\mathit{n}_{\mathit{k}}/\partial\mathit{t}$ its rate of change due to collisions.  Approximation of $\mathit{n}_{\mathit{k}}$ by the Bose distribution $\overline{\mathit{n}}_{\mathit{k}}$ (equivalent to the assumption that deviations from equilibrium are small\cite{carruthers}) yields
\begin{equation}
\frac{\partial\mathit{n}_{\mathit{k}}}{\partial\mathit{t}}\approx\mathbf{c}_{\mathit{k}}\!\cdot\!\nabla\!\mathit{T}\frac{\hbar\omega_{\mathit{k}}}{\mathit{k}_\mathit{B}\!\mathit{T}^{2}}\frac{\mathit{e}^{\hbar\omega_{\mathit{k}}/\mathit{k}_{\mathit{B}}\!\mathit{T}}}{\left(\mathit{e}^{\hbar\omega_{\mathit{k}}/\mathit{k}_{\mathit{B}}\!\mathit{T}}-1\right)^{\!2}}.
\end{equation}  

Callaway was the first to distinguish between the different effects of normal (i.e. momentum-conserving) and Umklapp (i.e. non-momentum-conserving) scattering processes on the thermal conductivity.  He employed the characterization that the equilibrium state approached through normal processes reflects the displaced Bose-Einstein distribution\cite{klemens}
\begin{equation}
\mathit{n(\lambda)}=\frac{1}{\mathit{e}^{(\hbar\omega_{\mathit{k}}-\lambda\cdot\mathbf{k})/\mathit{k}_{\mathit{B}}\!\mathit{T}}-1},
\end{equation}
\noindent where $\overrightarrow{\lambda}(T) =\beta(T)\,\nabla\!\mathit{T}/\mathit{T}$ lies in the direction of the thermal gradient; $\beta(T)$ is determined below.  In contrast, Umklapp processes cause the mode occupation numbers to relax to the static equilibrium distribution $\overline{\mathit{n}}_{\mathit{k}}$.  Using the relaxation-time approximation\cite{carruthers}, which postulates exponential relaxation to equilibrium\cite{klemens}, Callaway expressed $\partial\mathit{n}_{\mathit{k}}/\partial\mathit{t}$ as 
\begin{equation}
\frac{\partial\mathit{n}_{\mathit{k}}}{\partial\mathit{t}} = \frac{\mathit{n(\lambda)}-\mathit{n}_{\mathit{k}}}{\tau_{\mathit{n}}}+\frac{\overline{\mathit{n}}_{\mathit{k}}-\mathit{n}_{\mathit{k}}}{\tau_{\mathit{u}}},
\end{equation} 
\noindent where $\tau_{\mathit{n}}$ represents excitation lifetimes corresponding to normal processes and $\tau_{\mathit{u}}$ those corresponding to Umklapp processes.  In Callaway's treatment, these are not simply added, as under heat flow the thermal distribution of the gas relaxes differently for the two types of scattering processes.  

Since $\lambda\cdot\mathbf{k}$ is small, $\mathit{n(\lambda)}$ can be approximated to first order as
\begin{equation}
\mathit{n(\lambda)}\approx\overline{\mathit{n}}_{\mathit{k}}+\frac{\lambda\cdot\mathbf{k}}{\mathit{k}_{\mathit{B}}\!\mathit{T}}\frac{\mathit{e}^{\hbar\omega_{\mathit{k}}/\mathit{k}_{\mathit{B}}\!\mathit{T}}}{\left(\mathit{e}^{\hbar\omega_{\mathit{k}}/\mathit{k}_{\mathit{B}}\!\mathit{T}}-1\right)^{\!2}}.
\end{equation}
The deviations $\delta\mathit{n}_{\mathit{k}}=\mathit{n}_{\mathit{k}}-\overline{\mathit{n}}_{\mathit{k}}$ can then be calculated from (5), using (3) and (6):
\begin{equation}
\delta\mathit{n}_{\mathit{k}}=\left(-\tau_{\mathit{c}}\mathbf{c}_{\mathit{k}}+\frac{\tau_{\mathit{c}}\beta}{\tau_{\mathit{n}}\hbar\omega_{\mathit{k}}}\mathbf{k}\right)\cdot\nabla\!\mathit{T}\frac{\hbar\omega_{\mathit{k}}}{\mathit{k}_\mathit{B}\!\mathit{T}^{2}}\frac{\mathit{e}^{\hbar\omega_{\mathit{k}}/\mathit{k}_{\mathit{B}}\!\mathit{T}}}{\left(\mathit{e}^{\hbar\omega_{\mathit{k}}/\mathit{k}_{\mathit{B}}\!\mathit{T}}-1\right)^{\!2}},
\end{equation}			
\noindent where $\tau^{-1}_{\mathit{c}}=\tau^{-1}_{\mathit{n}}+\tau^{-1}_{\mathit{u}}$\cite{betanote}.

The magnitude of $\lambda(T)$, and thus of $\beta(T)$, is determined by applying the condition that the normal collisional processes conserve momentum, namely that
\begin{equation}
\sum_{\mathit{k}}\left(\frac{\partial\mathit{n}_{\mathit{k}}}{\partial\mathit{t}}\right)_{\mathit{\!\!\!n}}\mathbf{k}=\sum_{\mathit{k}}\frac{\mathit{n(\lambda)}-\mathit{n}_{\mathit{k}}}{\tau_{\mathit{n}}}\,\mathbf{k}=0,
\end{equation}					
\noindent which through application of (5), (3), and (7) yields the expression
\begin{equation}
\beta(T)=-\frac{\sum_{\mathit{k}}(\tau_{\mathit{c}}/\tau_{\mathit{n}})(\mathbf{c}_{\mathit{k}}\cdot\nabla\!\mathit{T})\mathit{C}_{\mathit{k}}\mathbf{k}/(\hbar\omega_{\mathit{k}})}{\sum_{\mathit{k}}(1/\tau_{\mathit{n}})(1-\tau_{\mathit{c}}/\tau_{\mathit{n}})(\mathbf{k}\cdot\nabla\!\mathit{T})\mathit{C}_{\mathit{k}}\mathbf{k}/(\hbar\omega_{\mathit{k}})^{2}},
\end{equation}		
\noindent where 
\[\mathit{C}_{\mathit{k}}=\frac{(\hbar\omega_{\mathit{k}})^{2}}{\mathit{k}_\mathit{B}\!\mathit{T}^{2}}\frac{\mathit{e}^{\hbar\omega_{\mathit{k}}/\mathit{k}_{\mathit{B}}\!\mathit{T}}}{\left(\mathit{e}^{\hbar\omega_{\mathit{k}}/\mathit{k}_{\mathit{B}}\!\mathit{T}}-1\right)^{\!2}}\] 
\noindent is the heat capacity of a given spin-wave mode.  Insertion of (7) and (9) into (1) yields $\kappa$.   

The thermal conductivity calculated using the well-known single-mode relaxation time (SMRT) approximation\cite{klemens} corresponds to $\beta = 0$, as in this approach normal processes are assumed to behave like Umklapp ones and relax to the (undisplaced) Bose distribution.  In Callaway's approach, normal processes, since they relax to a displaced equilibrium distribution, in fact contribute to thermal drift: a heat current flows even in the absence of a temperature gradient\cite{klemens}.

In order to use (7), (9), and (1) to calculate the heat current, the lifetimes (inverse linewidths) $\tau_{\mathit{n}}$ and $\tau_{\mathit{u}}$ corresponding to each relevant scattering process must be known.  Previous applications of Callaway's theory in calculations of the thermal conductivity relied on the use of adjustable parameters in expressions for $\tau_{\mathit{n}}$ and $\tau_{\mathit{u}}$ for each collisional process thought to occur\cite{callaway,asen-palmer}, as these were not known independently.  

\begin{figure}
\includegraphics[keepaspectratio=true,clip=true,trim=44mm 106mm 60mm 75mm,width=86mm]{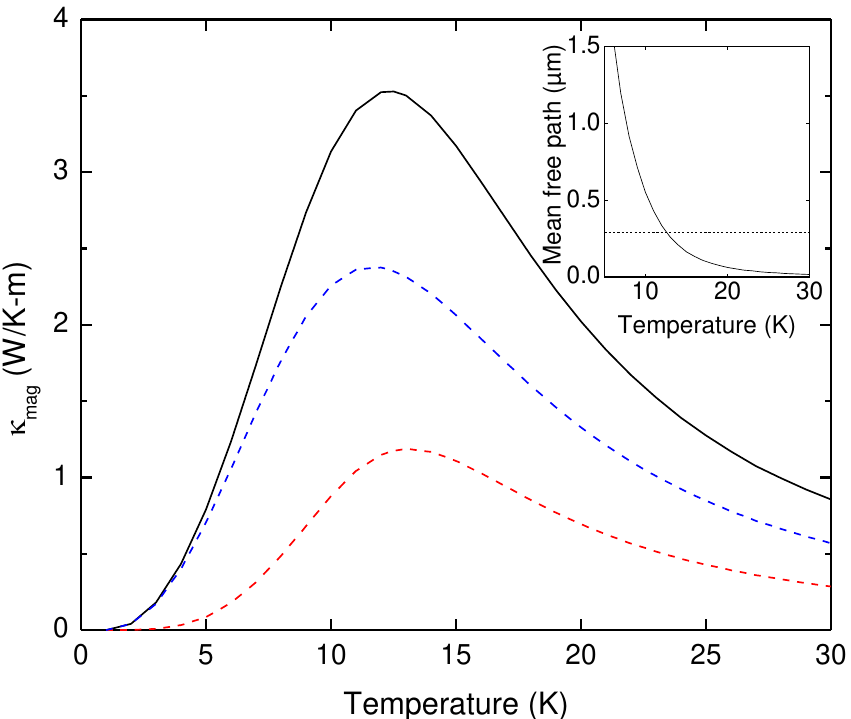}
\caption{\label{1} The magnon thermal conductivity of Rb$_{2}$MnF$_{4}$ (black line), computed as described in the text.  The growing magnitude up to $\sim\!15$ K of the heat current deriving, in Callaway's formulation, from the assumption of a displaced equilibrium distribution (dashed red line), relative to that corresponding to the SMRT approximation (dashed blue line) demonstrates the increasing importance of transport by the drift current with increasing temperature.  Inset: the mean free path in the interacting magnon gas (solid line) and the mean distance $\mathit{L}/2$ (dashed line) traveled by a magnon before colliding with an intracrystalline boundary with average spacing $\mathit{L} = 0.58 \pm 0.04$ $\mu$m\cite{bayrakcipreprint}.}
\end{figure}

For certain types of microscopic collisional processes, including scattering between phonons and scattering between magnons, the respective contributions of $\tau_{\mathit{n}}$ and $\tau_{\mathit{u}}$ to the combined lifetime $\tau_{\mathit{c}}$ cannot be distinguished by means of experiment.  However, in a numerical evaluation of the linewidths corresponding to a given process, it is straightforward to identify the contribution of each component to the overall linewidth, simply by keeping track of how momentum conservation is achieved in each individual scattering event that contributes to the linewidth.   In contrast, scattering of phonons or magnons from defects and boundaries is taken to contribute to the Umklapp component of the linewidth, as momentum is not conserved in such processes.  Such scattering can often be extracted experimentally. 

In recent neutron spin-echo experiments on the 2D antiferromagnet Rb$_{2}$MnF$_{4}$, the combination of 4-magnon (two-in/two-out) scattering processes and boundary scattering was shown to agree with the magnon linewidth data\cite{bayrakcipreprint}.  Thanks to this excellent description, which was achieved without the use of adjustable parameters, numerical evaluation can be used to separate $\tau_{\mathit{n}}$ and $\tau_{\mathit{u}}$ for the 4-magnon scattering for the purpose of calculating $\kappa_{\mathrm{mag}}$ within Callaway's framework.  The boundary scattering was taken from the lowest-temperature data (at 3 K)\cite{bayrakcipreprint}.  The main panel of Fig. 1 shows the resulting $\kappa_{\mathrm{mag}}$ calculated for Rb$_{2}$MnF$_{4}$.  This represents the first application of Callaway's theory without the use of any adjustable parameters. 

We also calculated the mean free path determined by the 4-magnon (two-in/two-out) interactions and the dispersion relation (Fig. 1, inset).  Below $\sim\!12$ K, the mean free path is longer than the mean distance traveled to an intracrystalline boundary or defect, and the thermal conduction is thus determined by ballistically-transported spin-waves that collide with boundaries or defects, where they are absorbed and emitted. At higher temperatures, the conductivity is increasingly limited by magnon-magnon interactions within the magnon gas, and the boundaries play a less significant role.

A qualitatively similar broad peak is also observed in the in-plane magnon-mediated thermal conductivity inferred from measurements in several square-lattice antiferromagnetic cuprates\cite{hess,hofmann,berggold}.  We used the above method to calculate the thermal conductivity in the cuprate La$_{2}$CuO$_{4}$, which is isostructural to Rb$_{2}$MnF$_{4}$.  Using the numerical methods described in Ref. 10\cite{bayrakcipreprint}, we calculated the linewidth for 4-magnon (two-in/two-out) scattering with $\mathit{S} = 1/2$ over the full Brillouin zone, using the low-temperature value for the nearest-neighbor exchange constant $\mathit{J}$ in La$_{2}$CuO$_{4}$.  We assumed that the renormalization of the exchange constant with temperature in La$_{2}$CuO$_{4}$ follows that measured in Rb$_{2}$MnF$_{4}$ (adjusted appropriately by $\mathit{S}$ and $\mathit{J}$ to reflect the different temperature scales).  Because the magnetic anisotropy is considerably smaller relative to the exchange constant in La$_{2}$CuO$_{4}$ than in Rb$_{2}$MnF$_{4}$, for calculational purposes we treated La$_{2}$CuO$_{4}$ as if its dispersion were gapless; the calculated linewidth results thus represent a slight overestimate.  In these numerical evaluations, iterations were performed until the linewidths converged\cite{bayrakcipreprint}; this convergence was 0.6\% or better over the full Brillouin zone for temperatures between 270 and 680 K.  We took the boundary spacing $\mathit{L}$ to be the temperature-independent mean free path of $558 \pm 140$ $\mathrm{\AA}$ that was fitted between 70 and 158 K in Ref. 5\cite{hess} and used this as the single adjustable parameter in calculating the component of the linewidth due to boundary scattering, as described in Ref. 10\cite{bayrakcipreprint}.  

\begin{figure}
\includegraphics[keepaspectratio=true,clip=true,trim=42mm 92mm 59mm 88mm,width=86mm]{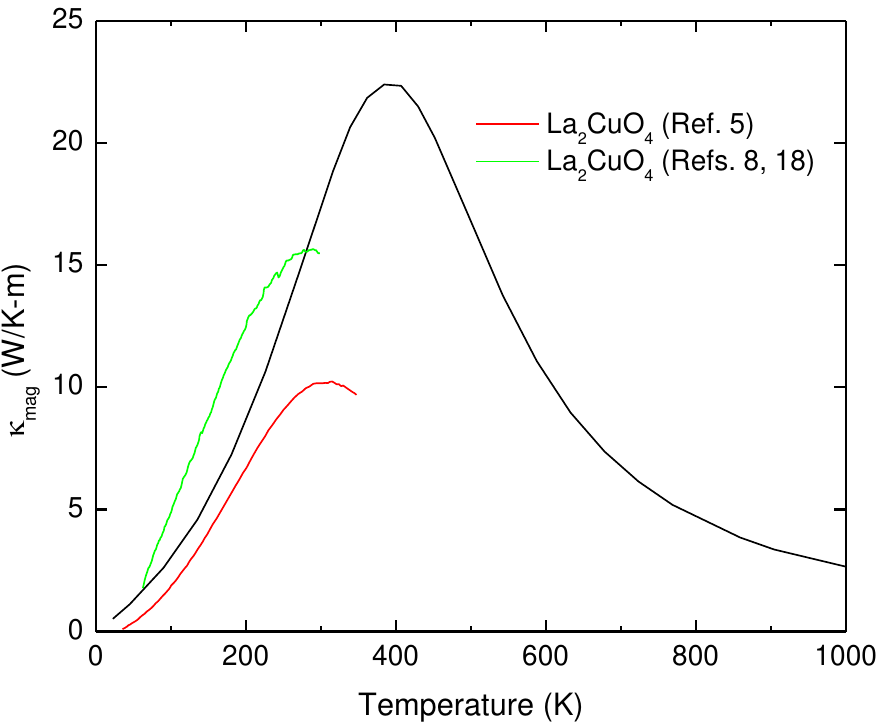}
\caption{\label{2} Extracted and calculated values of $\kappa_{\mathrm{mag}}$ in La$_{2}$CuO$_{4}$.  The extracted values were determined by subtracting the estimated phonon contribution from the measured experimental data for $\kappa$.}
\end{figure}

In Fig. 2, we compare these calculated results for $\kappa_{\mathrm{mag}}$ in La$_{2}$CuO$_{4}$ with representative extracted experimental results\cite{hess,berggold,yan}.  The agreement is reasonable, considering the strong dependence of the calculated results on the boundary scattering (which was not measured directly), and in light of the large variation between samples\cite{hess,berggold,yan}.  Deviations in La$_{2}$CuO$_{4}$ from a simple Heisenberg model, including ring exchange\cite{coldea}, may also be relevant.  Furthermore, the calculated peak in La$_{2}$CuO$_{4}$ falls in the vicinity of $\mathit{T}_{\mathit{N}}$, which is roughly 325 K.  At such high temperatures, the assumptions underlying the theory begin to break down, and other effects may become important.  Finally, there are experimental uncertainties in the separation of the magnon-mediated and phononic contributions to the thermal conductivity. 

High-resolution magnon linewidth data over the full Brillouin zone thus establish a detailed correspondence between microscopic interactions (the magnon decay processes) and a macroscopic transport property (the bulk thermal conductivity).  In order to evaluate the thermal conductivity from the linewidths of the excitation in question, quantitative description of the linewidths by theory is essential, as the contributions of normal and Umklapp processes cannot be separated experimentally.  In Rb$_{2}$MnF$_{4}$, such agreement with theory was found (using no free parameters), and this agreement makes it possible to link the magnon spectral function to the transport.  The same approach can also be employed in phonon systems, for which theoretical descriptions of the thermal conductivity have heretofore involved multiple adjustable parameters\cite{asen-palmer}.  

This formalism provides a complete description of the thermal conductivity over a very wide temperature range. If the magnon (or phonon) dispersion relations are known, this description involves only a single unknown parameter, the domain boundary (or dislocation) distance $\mathit{L}$.  This quantity can thus be extracted reliably from existing thermal conductivity data. This should provide a greatly improved basis for efforts to use the thermal conductivity as a probe of correlated metals and superconductors\cite{foo,doiron} and to design nanostructured materials with thermal properties optimized for energy conversion and other applications\cite{hochbaum}.


%


\begin{acknowledgments}
The work in Stuttgart was supported by the German Science Foundation under grant SFB/TRR 80.  
\end{acknowledgments}


\begin{thebibliography}{19}
\bibitem [{bet()}]{betanote}%
  \BibitemOpen
  \href@noop {} {}\bibinfo {note} {Our definition of $\beta$ differs from that
  used in Ref. 14 because of our allowing for nonlinearity of the excitation
  dispersions. The two are related by $\beta \mathbf{k} =
  -\hbar\omega_{\mathit{k}}\beta_{\mathrm{Callaway}} \mathbf{c}_{\mathit{k}}$.
  Eq. 6 in Ref. 14 matches our result for $\delta\mathit{n}_{\mathit{k}}$ if
  Callaway's $\beta$ is replaced by ours, using the above
  relation.}\BibitemShut {Stop}%
\end{thebibliography}
%
\providecommand{\noopsort}[1]{}\providecommand{\singleletter}[1]{#1}%

\end{document}